\documentstyle[11pt,aasms]{article}

\tighten

\begin{document}



\def\pmb#1{\setbox0=\hbox{#1}%
  \kern0.00em\copy0\kern-\wd0
  \kern0.03em\copy0\kern-\wd0
  \kern0.00em\raise.04em\copy0\kern-\wd0
  \kern0.03em\raise.04em\copy0\kern-\wd0\box0 }

\def\pp{\parshape=2 -0.25truein 6.75truein 0.5truein 6truein}

\def\ref #1;#2;#3;#4;#5{\par\pp #1 #2, #3, #4, #5}
\def\book #1;#2;#3{\par\pp #1 #2, #3}
\def\rep #1;#2;#3{\par\pp #1 #2, #3}

\def\undertext#1{$\underline{\smash{\hbox{#1}}}$}
\def\simlt{\lower.5ex\hbox{$\; \buildrel < \over \sim \;$}}
\def\simgt{\lower.5ex\hbox{$\; \buildrel > \over \sim \;$}}
\def\la{\lower.5ex\hbox{$\; \buildrel < \over \sim \;$}}
\def\ga{\lower.5ex\hbox{$\; \buildrel > \over \sim \;$}}
\def\r{\hangindent=1pc \noindent}
\def\j21{J_{-21}}
\def\lyal {{\rm Ly} \alpha}
\def\omigm {\Omega_{\scriptscriptstyle\rm IGM}}
\def\nhbr {N_{\scriptscriptstyle\rm HI,br}}
\def\mj {M_{\scriptscriptstyle\rm J}}
\def \tauhely{\tau ^{\scriptscriptstyle\rm He \,II}
_{\scriptscriptstyle Ly \alpha}}
\def \tauhegp{\tau ^{\scriptscriptstyle\rm He \,II}
_{\scriptscriptstyle GP}}
\def \tauhgp{\tau ^{\scriptscriptstyle\rm H \,I}
_{\rm \scriptscriptstyle GP}}
\def \nh{N _{\scriptscriptstyle\rm H \,I}}
\def \nhe{N _{\scriptscriptstyle\rm He \,II}}
\def \whmin{W^{\scriptscriptstyle\rm H \, I} _{
\scriptscriptstyle\rm min}}
\def \ang{{\rm \AA}}
\def\sc{\scriptscriptstyle}

\def\etal{{et~al.}}
\def\noi{\noindent}
\def\bs{\bigskip}
\def\ms{\medskip}
\def\ss{\smallskip}
\def\ob{\obeylines}
\def\l{\line}
\def\hrf{\hrulefill}
\def\hf{\hfil}
\def\q{\quad}
\def\qq{\qquad}
\renewcommand{\deg}{$^{\circ}$}
\newcommand{\um}{$\mu$m}
\newcommand{\uk}{$\mu$K}
\newcommand{\qrms}{$Q_{rms-PS}$}
\newcommand{\n}{$n$}
\newcommand{\cdmr}{${\bf c}_{\rm DMR}$}
\newcommand{\xrms}{$\otimes_{RMS}$}
\newcommand{\gt}{$>$}
\newcommand{\lt}{$<$}
\newcommand{\ldl}{$< \delta <$}


\bs
\bs
\bs
\bs
\bs
\bs
\bs
\bs
\title{On The Source of Ionization of The Intergalactic Medium at~$z \sim 2.4$}

\author{Shiv K. Sethi and Biman B. Nath 
}
\noindent
\affil{Inter-University Centre for Astronomy \& Astrophysics,
Post Bag 4, Pune 411007, India}

\bs
\bs
\bs

\begin{abstract}

We use the recent detection of He II absorption at $z=2.2\hbox{--}2.6$
in the line of sight of the quasar HS1700+64 to put bounds on
the sources of ionization. We find that given the uncertainty in
$\tauhgp$ and the model of absorption in the intergalactic medium (IGM),
a wide range of possible
sources of ionization is still allowed by the observations. 
We show that a significant contribution from star forming galaxies
is allowed and is consistent with the proximity effect.
In the case of photoionization by the UV background radiation, 
the contribution
of star forming galaxies to the intensity at $912 \> {\rm \AA}$ can
be within a range of $\sim 1 \hbox{--} 15$ times that of the
quasars. We also investigate the case of a collisionally
ionized intergalactic medium. We show that although collisional
ionization can be a dominant source of ionization at $z \sim 2.4$
(taking into account the upper limit on Compton y-parameter of the
microwave background radiation), the thermal state of the IGM cannot
yet be determined. We also consider Sciama's model of radiatively
decaying neutrinos and 
show that the model of decaying neutrino is consistent
with the observations.

\end{abstract}

\keywords{intergalactic medium  - quasars :
absorption lines - cosmology :
miscellaneous}

\section{INTRODUCTION}
Absorption studies in the line of sight of high redshift quasars
are important sources of our knowledge of the density and state of 
ionization of the intergalactic medium. Among various tests, Gunn-Peterson 
(GP) test, looking for absorption troughs in the
quasar spectra shortward of $\lyal$ emission line is probably the
most important one. The lack of any such absorption trough implies
that hydrogen
in the IGM has been in a highly ionized state since redshifts as high
as $z \sim 5$. The spectra are, however, crowded with numerous 
discrete absorption lines ($\lyal$ lines) showing the inhomogeneity in the IGM.
The existence of a UV background radiation at high redshift
is also inferred from the proximity effect,---the change in the 
number density of $\lyal$ lines in the
vicinity of quasars (Bajtlik, Duncan, \& Ostriker 1988).
The IGM is usually thought to have been photoionized by the UV
background radiation, whose intensity at the Lyman limit (for
H I, at $912 \> {\rm \AA}$) has been estimated from the proximity
effect. Observations show that the intensity
of the background radiation at $912 \> {\rm \AA}$ in the redshift
range of $z \sim 2\hbox{--}3$ is $J_{912,-21} \sim 10^{\pm 0.5}$, in the
standard unit of $10^{-21}$ erg cm$^{-2}$ s$^{-1}$ sr$^{-1}$ Hz$^{-1}$
(Bechtold 1994; Fernandez-Soto \etal\ 1995). 

The ratio of opacities due to various elements with different
ionization thresholds would give us crucial information about the
intensity and the spectrum of the ionizing background as well as
the thermal state of the IGM.
Studies of absorption by He II (singly ionized helium) atoms have
only been possible with the launch of UV telescopes in space.
Jakobsen \etal (1994) detected the absorption due to He II atoms
for the first time, but could only put lower bounds on the opacity
(of $\tau \ga 1.7$ at $z \sim 3.3$ in the line of sight of Q0302-003).
>From this lower limit, they and others (Madau and Meiksin 1994)
put bounds on the shape of the spectrum of the ionizing background.
Part of the opacity is possibly due to the He II atoms associated with
$\lyal$ lines (from line blanketing), and part of it is from absorption
in the diffuse IGM (the He II Gunn-Peterson effect). If one denotes the 
ratio of the intensity of the background radiation at $912 \> {\rm \AA}$ 
and $228 \> {\rm \AA}$ (the Lyman limit for H I and He II) as $S_L= 
J_{912}/ J_{228}$, then it was shown that $S_L \ga 40$, so that the 
observed lower bound on He~II opacity could be explained by a combination 
of line blanketing and He~II Gunn-Peterson effect.

Madau and Meiksin (1994) showed that such a soft ionizing radiation
can be attributed to optically detected quasars, after allowing for
continuum absorption by the intervening $\lyal$ absorbers, or, to a 
combination of radiation from quasars and star forming galaxies 
(Miralda-Escud\`e and Ostriker 1990). The degree of softness
would depend on the combination, and on the degree of
absorption. With only a lower limit on $S_L$, however, one is still left 
with a number of possible combinations.

Recently, Davidsen \etal (1996) have detected an unambiguous trough
in the spectrum of HS1700+64, and estimated the opacity due to
He II atoms to be $\tau \sim 1 \pm 0.07$ in the redshift range
$z \sim 2.2-2.6$. We argue below that this result allows one
to put both upper and lower limits on $S_L$,---in the case when all
of the opacity comes from line blanketing and in the case of partial
contribution from He II Gunn-Peterson effet, respectively. We use
these bounds to draw conclusions on the allowed combinations
of ionizing radiation from quasars, star forming galaxies and radiatively 
decaying
neutrinos (Sciama 1990). In addition, we discuss the implications of 
this observation on the thermal state 
of the IGM, since a contribution from collisional ionization would also
change the ratio of observed values of H I and He II opacity.

The structure of the paper is as follows. \S 2 calculates the bounds
on the softness of the UV background radiation, or more generally,
the ratio of the GP opacities for H~I and He~II, using the He~II
opacity. We then discuss the implications for various sources of
ionization in \S 3. Finally, we summarise the implications of our
results in \S 4.

\section{Bounds on the softness parameter}

The observed He II opacity is partly due to the He II atoms in the 
$\lyal$ lines, {\it i.e.}, due to line blanketing. This contribution
to the total opacity can be estimated using a distribution for the
$\lyal$ lines, and using the fact that the H I and He II column densities
are related as $\nhe =1.8 \> S_L \> \nh$. This relation assumes 
photoionization equilibrium of the gas with the UV background radiation
(whose softness parameter is $S_L$). For the $\lyal$ lines, we will use the 
rest equivalent width distribution
(Murdoch \etal 1986) (as in Madau and Meiksin (1994)),
\begin{eqnarray}
{ \partial ^2 N \over \partial W \partial z}&=& 40.7 \exp \Bigl (-{W^{\scriptscriptstyle\rm H \,I} \over
W_{\ast}} \Bigr ) (1+z)^{2.46} \nonumber\\
&\qquad&\qquad (0.2 < W^{\scriptscriptstyle\rm
H \,I} < 2 \AA) \,; \nonumber\\
&=&11.4 \Bigl ({W^{\scriptscriptstyle\rm H \,I} \over W_{\ast}}
\Bigr )^{-1.5} (1+z)^{2.46} \nonumber\\
&\qquad&\quad
(W^{\scriptscriptstyle\rm H \,I} _{min} <
W^{\scriptscriptstyle\rm H \,I} < 0.2 \AA) \,,
\end{eqnarray}
where $W_{\ast}=0.3 \AA$. The opacity due to line blanketing is then
given by (Paresce, McKee
and Bowyer 1980),
\begin{equation}
\tau_{\it eff}(z)= {1+z \over \lambda _ \alpha} \int ^{W_{max}} _{W_{min}}
{\partial ^2 N \over \partial W \partial z} \,W \,dW
\end{equation}
where $\lambda _{\alpha}$ is the rest wavelength, and $\partial ^2 N
/ \partial W \partial z$ is the rest wavelength distribution of the
absorbers. We will also use a constant velocity parameter of
$b^{\scriptscriptstyle\rm H \,I}=35$
km/s, as in Madau and Meiksin (1994). This is not very unrealistic, since the
observed distribution is fairly peaked around this value of $b$
(see, e.g., Giallongo et al. 1996). For the case where the velocity of the atoms is dominated
by hydrodynamical motions, the velocity parameters of He II and H I are
related as $b^{\scriptscriptstyle\rm He \,II} = \xi \> 
b^{\scriptscriptstyle\rm H \,I}$, where $\xi=1$. When thermal motions 
dominates, then $\xi=0.5$.

Using the above distribution, the H I opacity due to lines at a redshift
$z \sim 2.4$ is estimated as $\tau ^{\scriptscriptstyle\rm H \, I}
_{\scriptscriptstyle Ly \alpha} =0.235$, for a lower limit of $W_{min}
=0.01 \> {\rm \AA}$. (The value is not very sensitive  to the lower limit
of $W$.) The flux decrement shortward of $\lyal$ line for HS1700+64
as observed by Davidsen \etal (1996) corresponds to $\tau \sim 0.22$.
Obviously, using the average distribution above to calculate the opacity
is not accurate to less than $10 \%$. Moreover, since the
two opacities (from observation and what is expected from line blanketing)
are close, it shows that the corresponding $\tauhgp$ is a small number.
It is worth recalling here that Steidel and Sargent (1987) put an upper
limit on $\tauhgp $ at $z \sim 2.5$ of $\la 0.05$.

In the case where the observed He II opacity stems only from line 
blanketing, the relation between $S_L$ and the He II opacity is shown
in solid lines in Figs.~1a and~1b. The horizontal lines represent the
He~II opacity observed by Davidsen \etal (1996). Figs.~1a and~1b correspond
to two values of the minimum H I equivalent width of $\lyal$ lines,
$\whmin =0.0025 \> {\rm \AA}$ and $\whmin=0.01 \> {\rm \AA}$. The latter
value is for H I column density $\sim 2 \times 10^{12}$ cm$^{-2}$. 
It is seen that with $\whmin=0.01 \> {\rm \AA}$,
for velocity broadened lines
($\xi=1$, $S_L \sim 89 \pm 21$, and for thermal broadening ($\xi =0.5$),
$ 512 \pm 138 $. These are the upper limits on the value of $S_L$, for
the He II opacity of $1 \pm 0.07$ (Davidsen \etal 1996).
The values of $S_L$ are lowered, when one adds the possible
contribution from the opacity due to diffuse medium. The opacities
of He II and H I due to the diffuse IGM are related as $\tauhegp
\sim 0.45 \> S_L \> \tauhgp$. We will use a value of $\tauhgp \sim
0.05$, corresponding to the upper limit put by Steidel and Sargent (1987).
Lower values of $\tauhgp$ will give larger values of $S_L$. The dashed lines
in Fig. 1 show the dependence of He II opacity on $S_L$. For velocity
(thermal) broadened lines, we get $S_L \sim 18 \pm 2$  ($S_L \sim 25 \pm 2$).
These are the lower limits on $S_L$.

Extending the $\lyal$ distribution below $\nh \sim 2 \times 10^{12}$ 
cm$^{-2}$ down to $\whmin =0.0025 \> {\rm \AA}$ increases the He II opacity
for a given value of $S_L$, and therefore decreases the bounds on $S_L$
for a given opacity. With only line blanketing, for velocity broadened
lines, we find $S_L \sim 42 \pm 7$ and for thermal broadened lines, 
$S_L \sim 120 \pm 29$. Including a $\tauhgp =0.05$, we find, for
velocity broadening, $S_L \sim 16 \pm 2$ and for thermal broadening, 
$S_L \sim 21 \pm 2$.

In a high resolution study along the line of sight of Q1331+170, Kulkarni
\etal (1996) showed that the equivalent width distribution might be steeper
than the distribution in eqn (1). If confirmed, this will shift the
curves in Fig.~1 towards larger values of $S_L$.

\bs
\section{Bounds on sources of ionization}

\subsection{ Quasars}

Proximity effect---the change in the number density of $\lyal$ lines
in the vicinity of quasars---show that the intensity of the UV background
radiation at $z \sim 2.4$ at the Lyman limit ($912 \> {\rm \AA}$) is
$\j21 \sim 10^{ \pm 0.5}$ (Fernandez-Soto 1995). Part of the ionizing
radiation obviously comes from the quasars, which are the only known source
of UV photons at high redshifts, and several authors have
derived the intensity due to quasars (Miralda-Escud\`e and Ostriker 1990,
Madau 1992, Meiksin  \& Madau 1993, Haardt \& Madau 1996). We will follow the derivation by Madau 
(1992) using the
quasar luminosity function as given in Boyle (1991). 
(The luminosity function of quasars  at redshifts exceeding 3 remains an issue of controversy (for a recent discussion see  Pei 1995 and references therein) 
As in Madau (1992),
we will assume that the comoving number density of quasars is conserved
beyond $z \sim 3$ ( to be precise, $z_c$ of Boyle (1991)) till $z \sim 5$.
The spectrum of the quasars is similarly assumed to be $F_{\nu} \propto
\nu ^{-0.7}$, for $\lambda > 1216 \> \ang$, and $ F_{\nu} \propto \nu 
^{-1.5} $, for $\lambda < 1216 \> \ang$ (this value of spectral index is also 
favoured by recent observations (Tytler 1995)), as in Madau (1992).

The spectrum of the UV background radiation will be determined by the
intrinsic spectra of the quasars and absorption of the photons by the intervening
$\lyal$ absorption systems (the spectrum is also affected by the uniformly 
distributed hydrogen and helium, but this effect is negligible).  We use models 
A2 from Miralda-Escud\`e
and Ostriker (1990) and LA from  Meiksin and Madau (1993)  for the H I column density distribution of the
$\lyal$ lines. The second of these models (LA) is based on the fact that there 
exists a deficit of $\lyal$ systems with column densities $ \ga 10^{15} \rm \, cm^{-2}$. And so the real opacity of the universe at high redshifts might be lower than predicted
by A2. For a recent discussion of possible models of absorption by $\lyal$ 
clouds, see (Giroux and Shapiro 1996). Most allowed models of $\lyal$ 
absorption fall between models A2 and LA. We show in Fig. 2 the resulting spectrum of the UV background at $z \sim 2.4$
due to quasars. The strong break at  and $228 \> \ang$ is because
of absorption at the Lyman limit for He II atoms (for details, see  appendix). It
is seen that the value of $S_L$ of the ionizing background for quasars
alone is  $10\hbox{--}20$, depending on the absorption model.   We have neglected the 
contribution of recombination photons from the $\lyal$ systems to the
background flux. This contribution, as shown by  Haardt and Madau (1996) 
makes an insignificant difference to the value of $S_L$ at $z \simeq 2.4$.
We note here that the intensity at $912 \ang$ for radiation from only 
quasars falls short
of the inferred value from proximity effect.

Comparing with the bounds on $S_L$ from Fig. 1, it is seen that radiation
from only quasars is consistent with a value of $\tauhgp \simeq 0.05$.

\subsection {Star forming galaxies}

It is possible that young galaxies which are forming stars at high redshift 
emit a large number of UV photons. Several authors have discussed the effect of such star forming young galaxies on the background
radiation (Bechtold \etal 1987, Miralda-Escud\`e
and Ostriker 1990). The effect will depend on the intrinsic spectrum of the young
galaxies, their number density and evolution in redshift, and absorption by 
the intervening $\lyal$ absorption systems. Miralda-Escud\`e and Ostriker
(1990) used the UV spectrum from Bruzual (1983) (their Fig. 2), and we
also use the same analytical fit.

The intensity of the ionizing flux both on the history of galaxy
formation and the timescales of star formation in the young galaxies. We
use the model ``G2'' for galaxy formation and ``S1'' for star formation
history of Miralda-Escud\`e and Ostriker (1990). The model ``G2'' assumes
a Gaussian model of galaxy formation with a formation redshift of
$z_f=4.5$ with a span in redshift $w=2$. The normalisation of the
ionizing flux, which also includes the uncertainty of the absorption by
dust inside the galaxies, is then varied to change the
contribution of young galaxies to the ionizing flux. The ionizing flux
is most sensitive to the intrinsic spectrum of the young galaxies. For
a Bruzual-type intrinsic spectrum, the ionizing flux at redshift $z \sim 2.5$
receives contributions from photon emitted at redshifts $z \la 2.6\hbox{--}3$, 
depending on the absorption model and the normalisation. So, the
ionizing flux from young galaxies does not depend strongly on the
redshift evolution of galaxies, but only on the intrinsic spectrum
and the normalisation (relative to the quasars). We show typical
spectra of the ionizing flux from young galaxies and quasars, with
two models of absorptions discussed in the last section, in Fig. 3.

We show the effect of adding UV photons from young galaxies to that from
quasars in Fig. 4. It plots the value of $S_L$ as a function of the ratio
of intensities from young galaxies to that from quasars at $912 \> \ang$.
As expected, $S_L$ increases with the ratio, {\it i.e.}, the addition
of UV photons from young galaxies makes the spectrum of the background
radiation softer. This is because of two reasons. Firstly, the intrinsic
spectrum of star forming galaxies is softer than that of quasars. Secondly,
the absorption at the He II Lyman limit has a non-linear effect on the
softness parameter. The softer the spectrum is, the more abundant are the
singly ionized helium atoms in the $\lyal$ absorption systems, and the
more is the absorption, therefore, further increasing the softness parameter 
(Miralda-Escud\`e \& Ostriker 1990). The horizontal lines correspond 
to the values of $S_L$ inferred from the discussion in \S 2.

The bounds on the sources of photoionizing background radiation from
 Figs. 1 and 4 are summarised in Table 1. The column for A2 for the
case of $\tauhgp=0.05$ is left blank as the ratio $J_{912, YG}/ J_{912, QSO}$
is too small ($\ll 0.5$)
 to be relevant. It is seen from Table 1 that a wide ranging
values ($\sim 1\hbox{--}15$) of $J_{912, YG}/ J_{912, QSO}$ is possible
given the uncertainty in the absorption model and in the value of $\tauhgp$.
We note here that the total intensity is $\j21=1$ for $J_{912, \rm YG}/ J_{912, 
QSO}=6.7 ({\rm A2}), 4.1 ({\rm LA})$. This fact, and the values in Table 1,
imply that only values of $\tauhgp < 0.05$ are consistent with the proximity
effect results, especially for model A2.


\begin{table}
\centerline{Table 1: Allowed values of $J_{912,\rm YG}/ J_{912, \rm QSO}$ from $\tau _{\scriptscriptstyle \rm He II} (z=2.4)=1. \pm 0.07$ (Fig. 1).}
\centerline{`vel'
and `th' correspond to velocity and thermally broadened lines as in Fig. 1.}
\vskip 0.2in
\centerline{\begin{tabular}{lccc}
\tableline
\tableline
\multicolumn {1} {c}{} & \multicolumn {1}{c}{}&\multicolumn {2}{c}{$J_{912}(\rm YG)/J_{912}(QSO)$ for}\\
\multicolumn {2} {c}{}&
\multicolumn {1} {c}{A2}&\multicolumn {1} {c}{LA}\\
\tableline
Line blanketing & (vel)&$1.65 \pm 0.45$& $5.\pm 1.$\\
     & (th)&$9.5\pm 3.$&$14.7 \pm 2.25$\\
With $\tauhgp=0.05$ & (vel)& \hbox{--}   & $0.85 \pm 0.15$\\
      & (th)& \hbox{--}&$1.45 \pm 0.15$\\
\tableline
\tableline
\end{tabular}}
\end{table}

\subsection{ Collisional ionization}

It is also possible that the intergalactic medium is warm ($T_{IGM}
\sim 10^6$ K) and it is collisionally ionized to a large extent. Several
authors have pursued the models of reionizing the IGM by collisional
ionization, with various heating sources. For example, Tegmark, Silk and Evrard
1993 showed that galactic winds from low mass galaxies can heat the universe
to the extent that the IGM is collisionally ionized, and Nath \& Biermann
(1993) showed that cosmic rays from galaxies could also be a source of heating.
A warm IGM has also been invoked to inhibit the formation of
small scale structure in the universe (Blanchard \etal 1992). For
a temperature of the IGM  $ T_{IGM} \ga 10^5$ K, the hot electrons
will ionize H~I and He~II atoms to a varying degree, depending on the
temperature and the number density of the atoms. The ratio of the
Gunn-Peterson opacities of H~I and He~II atoms will therefore depend
on the temperature. The final effect on the opacities will depend both
on the temperature and the
softness of the photoionization background radiation which exists 
in any case.

We show in Fig. 5 the effect of adding collisional ionization.
We define $S_L ^{\it eff}$ as the ratio $\tauhegp / (0.45 \tauhgp)$ for the
case including collisional ionization. Note that for a photoionized IGM,
$S_L=\tauhegp / (0.45 \tauhgp)$. In other words, an IGM with both
collisional and photoionization mimics a photoionized IGM with the
inferred value of $S_L$ (from the observations of $\tauhegp$ and
$\tauhgp$) equal to $S_L ^{\it eff}$.
The value of $S_L ^{\it eff}$ first decreases at the ionization threshold
of He II, because In the absence of collisional ionization,
the intensity of $J$ at
$228 \> {\rm \AA}$ is small, with a low photoionization rate. Including
a high temperature makes collisional ionization more dominant and ionizes
He II abundantly. Collisional ionization does not make a large difference
at the ionization threshold of H I as the photoionization rate there is
already high. This explains the decrease of $S_L^{\it eff}$ at the ionization
threshold of He II in Fig. 5. The decrease is larger for the case of
UV radiation from quasars and young galaxies. This is because in the case of radiation from only quasars, the intensity of $J$ is larger at $228 \> {\rm \AA}$ than in the case of radiation from quasars and young galaxies 
(see Figures 2 and 3). This is why collisional ionization makes a large
difference in the latter case. The value of $S_L^{\it eff}$
 increases at higher temperatures when collisional ionization becomes 
the dominant source of ionization for hydrogen also.
At very high temperatures, the rates of collisional
ionization and recombination of both H I and He II have similar dependences
on the temperature (for the dependence of recombination coefficients on temperature, see Cen 1992). This explains the leveling off of the ratio at
high temperatures. 

We have assumed here that collisional ionization in the IGM does not
change the internal structures of the $\lyal$ absorption systems. It is
justified because the temperature of $\lyal$ absorption systems are
of the order $10^4\hbox{--}10^5 \> {\rm K}$, as inferred from the $b$ parameter
of the $\lyal$ lines, whose distribution has a peak around $35$ km/s (Giallongo et al. 1996).
This means that photoionization is the dominant process of ionization
inside the clouds.


We also calculate the limit on the temperature of the IGM from the COBE
limits on the Compton $y$ parameter, $y \la 1.5 \times 10^{-5}$ (Fixsen \etal\ 
1996). If the hot
electrons in the IGM loses energy to the microwave background photons
through inverse Compton scattering at $z \sim 2.4$, then the temperature
of the IGM is bounded as $T \la 1420 \,\rm eV$ (Fig. 5).

It is not yet possible to put limits on the temperature of the IGM
gas based on the analysis of $S_L ^{\it eff}$ , since the relative contributions
of quasars and young galaxies in the UV background radiation and the
value of $\j21$ are unknown.
However, if $\j21$ is determined in the future with better accuracy at
this redshift, and the value of $\tauhgp$ for this line of sight is
determined (in other words, the value of $S_L^{\it eff}$ is determined), then
such an analysis as described above will lead to limits on the temperature.

\subsection {Decaying neutrinos}

Another possible source of  photoionization of the IGM at high redshifts is 
radiatively decaying neutrinos (Sciama 1990; For details and relevant references
of Sciama's model see Sciama 1994b; Sethi 1996). In this paper, we consider 
the model of radiatively decaying neutrinos proposed by Sciama. In this model, 
massive  neutrinos of mass $ m_\nu\sim 30 \, \rm  eV$ decay into a massless neutrino
and a photon with a decay lifetime  $\tau \sim 10^{23} \, \rm sec$. Observation
of hydrogen-ionizing photon flux at the present epoch fixes the parameters $m_\nu$ and $\tau$ (Sciama 1995). We use $\tau = 2 \times 10^{23} \, \rm sec$
and $m_\nu = 27.4 \, \rm eV$, as in Sciama (1995). The resulting spectrum of the ionizing flux, at $z = 2.4 $, with 
the quasars and radiatively decaying neutrinos as the two sources is shown in
Fig 6. The value of $J_{-21}(912) $ is $\sim \{0.72,0.64 \}$ for the model of absorption \{LA, A2\}. Depending on the value of model of absorption, the resulting 
value of $S_L$ lies between $32$ and  $585$. These values are consistent
with those in Fig 1 and indicate a value of $\tauhgp < 0.05$. 

 The  feature that distinguishes
radiatively decaying neutrinos  from other soft sources of photoionization
like the young galaxies is the relative suppression of neutral helium-ionizing
flux. The observation of neutral helium resonant lines in 4 lyman-limit systems
 along the line of
sight to HS1700-64 can indicate the nature of ionizing background at that 
redshift (Reimers \& Vogel 1993; Sciama 1994a; Sethi 1996).   However, unlike the GP tests which probe
the ionization state of the IGM averaged over length scales $\sim \hbox{a few}
\times 10 \,\, \rm Mpc$, and  are therfore insensitive to the local sources of 
ionization, the state of ionization inside the lyman-limit systems could be 
caused by sources inside the cloud (Viegas \& Fria\c{c}a 1995). Therefore, for studying the
nature of background ionizing flux, we do not discuss the consequences of this
observation.

\subsection {He~I GP test}

HS1700+6416 has also been observed for He~I GP 
effect (Reimers \etal\ 1992); the reported upper limit on the number density of neutral helium $ n_{\scriptscriptstyle \rm HeI} \le 7
\times 10^{-12} \, \rm cm^{-3}$ (for $q_0 = 0$, $h_0 = 0.5$) translates to $\tau^{\scriptscriptstyle\rm He \,I}_{\scriptscriptstyle \rm GP} \la
0.05$ (for $q_0 = 0.5$, $h_0 = 0.5$). For photoionization models the GP optical
depth can be related to the background ionizing photons as:
\begin{equation}
\tau^{\scriptscriptstyle\rm He \,I}_{\scriptscriptstyle \rm GP}(z) =
{ 1.6 \times 10^{-3} h_0^3 \Omega_{\rm IGM}^2 (3+\alpha) (1+z)^{4.5} \over
 J_{-21}(504)} y_2.
\end{equation}
Here $\alpha \simeq 0.5$ is the local index of the spectrum of ionizing 
photons near the ionization threshold of He~I; $y_2 =  n_{\sc\rm HeII}/n_{\sc\rm
He}$ in the fraction of He~II.  For $\tauhegp \simeq 1$ at $z \simeq 2.7$, one
gets $y_2 = 5 \times 10^{-4}$ (for $\Omega_{\rm IGM} = 0.05$, $h_0 = 0.5$, $q_0 = 0.5$). Noting that $J_{-21}(504) \simeq 0.1 \hbox{--}1$ for all the models we
have studied (Figures (2), (3), and (6)) and inserting the 
value of $y_2$ in Eq. (3), one sees
that the He~I GP test is easily satisfied for HS1700+6416. (It should 
be pointed out that as different models of photoionization give quite different values for He~II ionizing
flux  (Figures (2), (3), and (6)), the fiducial 
values of $\Omega_{\rm IGM}$, $h_0$ and $q_0$ we 
use for the estimate given above will not satisfy He~II GP test
for all the models; however a 
scatter in these values does not change our conclusion).

\section{Summary and discussion}

The proximity effect suggests that $J_{-21}(912) \simeq 1$ at $z \simeq 2.4$. 
As is seen from Fig. 2, quasars contribute only a fraction of this flux.
If one assumes that the remaining flux is contributed by the star forming
galaxies, then, for model A2 (LA), one needs  $J_{912, \rm YG}/ J_{912, QSO}=6.7 (4.1)$. Fig. 4 gives the corresponding
values of the softness parameter. For A2 (LA), we find that $S_L=367 (68)$.

In Fig. 7 we show the value of $\tauhgp$ necessary to explain $\tau_{
\scriptscriptstyle \rm HeII} =1$, including line blanketing with
$\whmin=0.01 \> {\rm \AA}$, for velocity and thermally broadened lines.
It is seen that for the large values of $S_L$ implied by proximity effect,
the required $\tauhgp$ is very small. 
For $S_L = 68$, for thermally (velocity) broadened lines, $\tauhgp \sim 0.015 (0.002)$, which implies
$\Omega_{\rm IGM}h^{3/2} = 0.015(0.006)$. For $S_L=367$, for thermally
broadened lines, $\tauhgp \sim 0.0005$, implying  
$\Omega_{\rm IGM}h^{3/2} =0.002$. A note of caution is in order here.
We have used an average column density distribution of $\lyal$
lines to calculate the line blanketing opacity. To rely on 
 such small values  as  $\tauhgp \sim 0.0005$, however, one should use the
column density distribution of $\lyal$ lines in the particular line
of sight.

The fact that  $\tauhgp$ is not determined in the same lines of sight 
in which He II absorption has been detected 
is an obstacle in the study of the physical condition 
of the IGM. Although Davidsen \etal (1996) have determined the decrement
$D_A$ in the spectrum of HS1700+64, it is not enough for the purpose of 
studying the background radiation and the IGM gas. It is important to 
know the contribution from $\lyal$ lines and the diffuse IGM to this
decrement. Using an average distribution of $\lyal$ lines to determine this,
or using the value of $\tauhgp$ from other lines of sight, is not a useful
procedure. It is now thought that the IGM is probably not
homogeneous and there is a probability distribution for the value of
$\tauhgp$ at a given redshift (Reisenegger and Miralda-Escud\`e 1995).
We here note that the absence of a determination of $\tauhgp$ in the 
line of sight of
Q0302-003 also makes the study of the IGM in the vicinity of that
quasar and the UV background radiation much uncertain (Nath and Sethi 1995; 
Giroux, Fardal \& Shull 1995).


The quasar HS1700+64 has been studied in the past in various contexts.
Various elements in high ionization states (N~V and O~VI) were observed
in the Lyman limit systems along this line of sight (Vogel \& Reimer  1993).
If these states of ionization were to be attributed to the background
radiation it would imply a hard background source (with spectrum
like $\nu ^{-0.6}$) (Vogel \& Reimer  1993). Although it is not clear that these ionization
states are caused by background radiation, the consistency of such
an assumption with our analysis would require $\tauhgp > 0.05$ (Fig. 1).

A large contribution from young galaxies to the UV background radiation
will be accompanied by production of metals. According to the calculations
of Giroux and Shapiro (1996), the metallicity of the universe at $z \sim 3$
can be larger than that of Population II stars. However, their calculation
is based on continuous metal production. They have suggested that if 
metals were ejected from the galaxies, this problem will be alleviated
(see also Madau and Shull 1996).

To summarise, the observation of Davidsen \etal\ (1996) brings us 
closer to the source of ionization at $z=2.4$. Unlike the previous
studies based on the detection of He~II in Q0302-003, which permitted
only a lower limit on the softness parameter ($S_L$), this new observation
allows us to put an upper bound on $S_L$. Our study shows that this rules
out sources with very soft spectra (e.g., $S_L \ga 650$, see Fig. 1)
which is independent of the value of $\tauhgp$ in this line of sight.
However, as we have shown
in the previous sections, it is difficult to pinpoint the exact nature of
the ionizing source, because $S_L$ has a large range of allowed
values ($16 \la S_L \la 650$). 
Among other things, determination of $\tauhgp$
along this line of sight and high resolution observation of $\lyal$
lines would help resolve some of these issues.

\bigskip
We thank R. Srianand for several stimulating discussions and the
anonymous referee for his comments.

\newpage

\section{Figure captions}

\bs

\noindent
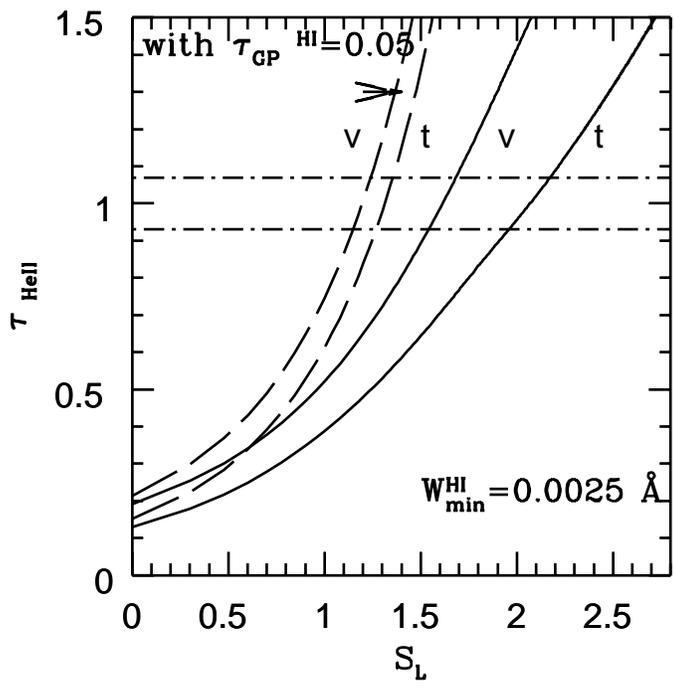
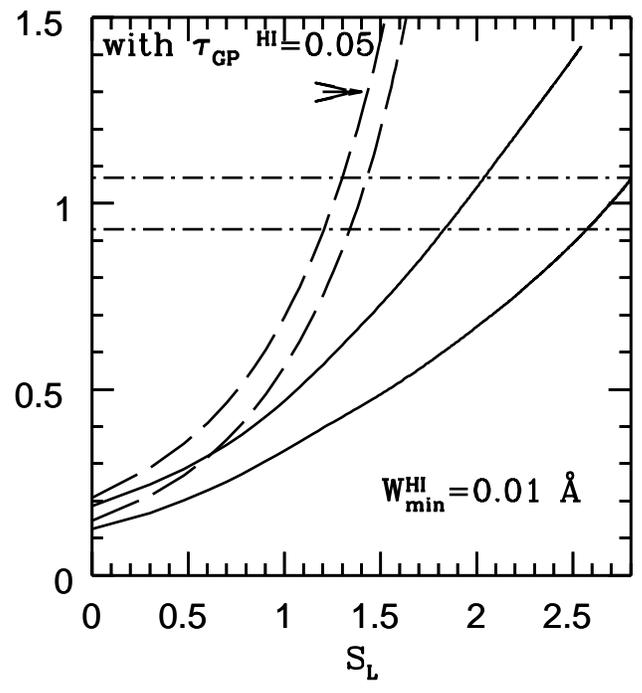
{\bf Figure 1} : We plot $\tau _{\scriptscriptstyle \rm He II}$ against
$S_L$ for two values of $\whmin=0.01, 0.0025$ $\ang$. Solid lines
show the opacity where only line blanketing is included, for velocity
and thermally broadened lines, which are labeled `v' and `t' 
respectively. Dashed lines show the corresponding opacity when
opacity from diffuse IGM is also included, for $\tauhgp=0.05$.
The horizontal lines correspond to $\tau _{\scriptscriptstyle \rm He II}
=1.0 \pm 0.07$, as observed by Davidsen \etal (1996).

\bs

\noindent
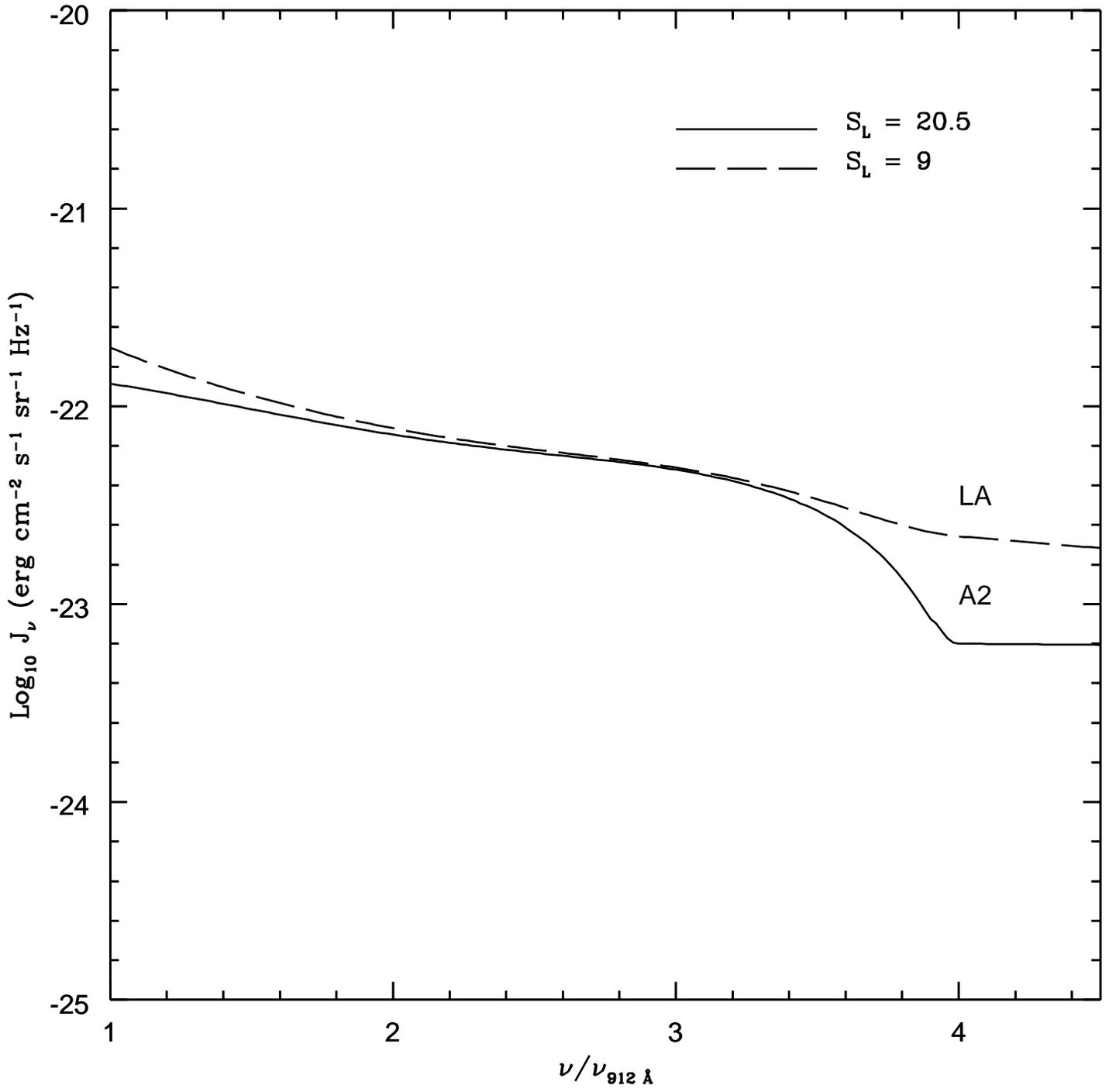
{\bf Figure 2} : The spectrum of the UV background radiation at
 $z =2.4$ due to
quasars is shown for two absorption models A2 and LA. The corresponding
values of the softness parameter $S_L$ are also shown.

\bs

\noindent
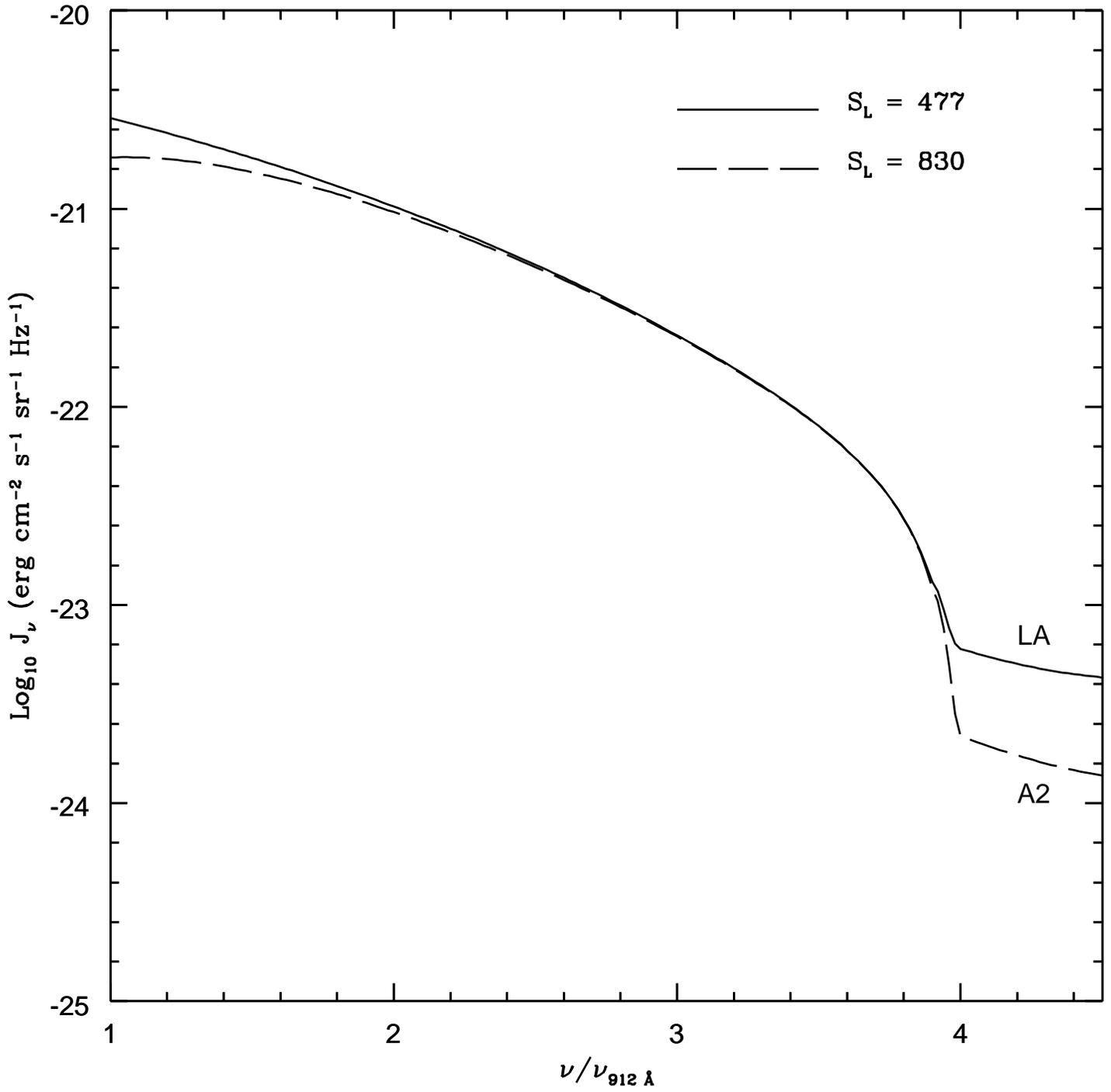
{\bf Figure 3} : A typical spectrum of the UV background radiation at $z=2.4$
due to quasars and young galaxies is shown for two absorption models
A2 and LA.
The corresponding
values of the softness parameter $S_L$ are also shown.

\noindent
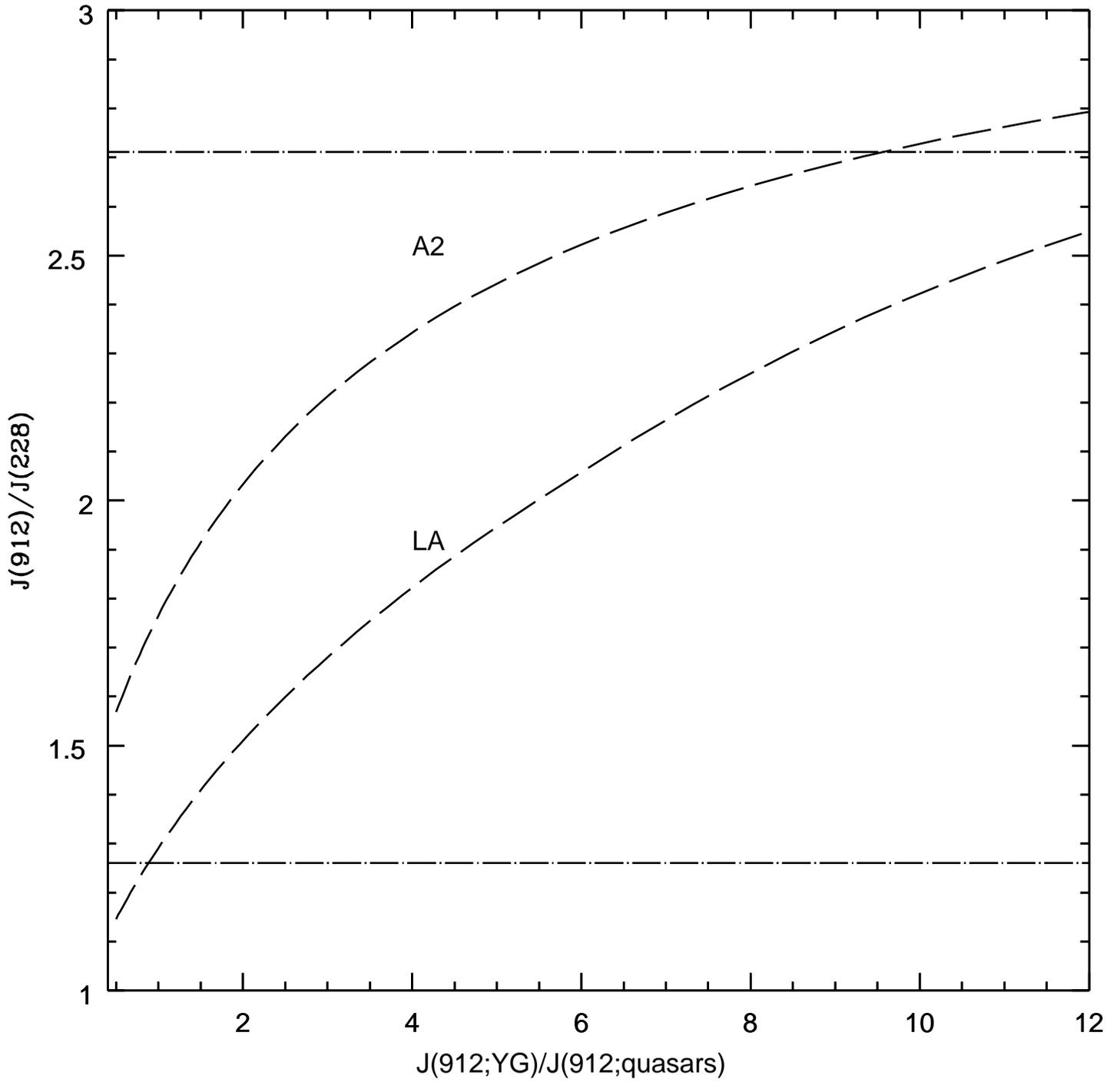
{\bf Figure 4} : $S_L$ is plotted against the ratio of contributions
to the UV background radiation from young galaxies to that from quasars, $J_{912, \rm YG}/ J_{912, 
QSO}$, for two absorption models A2 and LA. Constraints on the softness
parameter $S_L$ from \S 2
are shown as horizontal lines.

\bs
\noindent
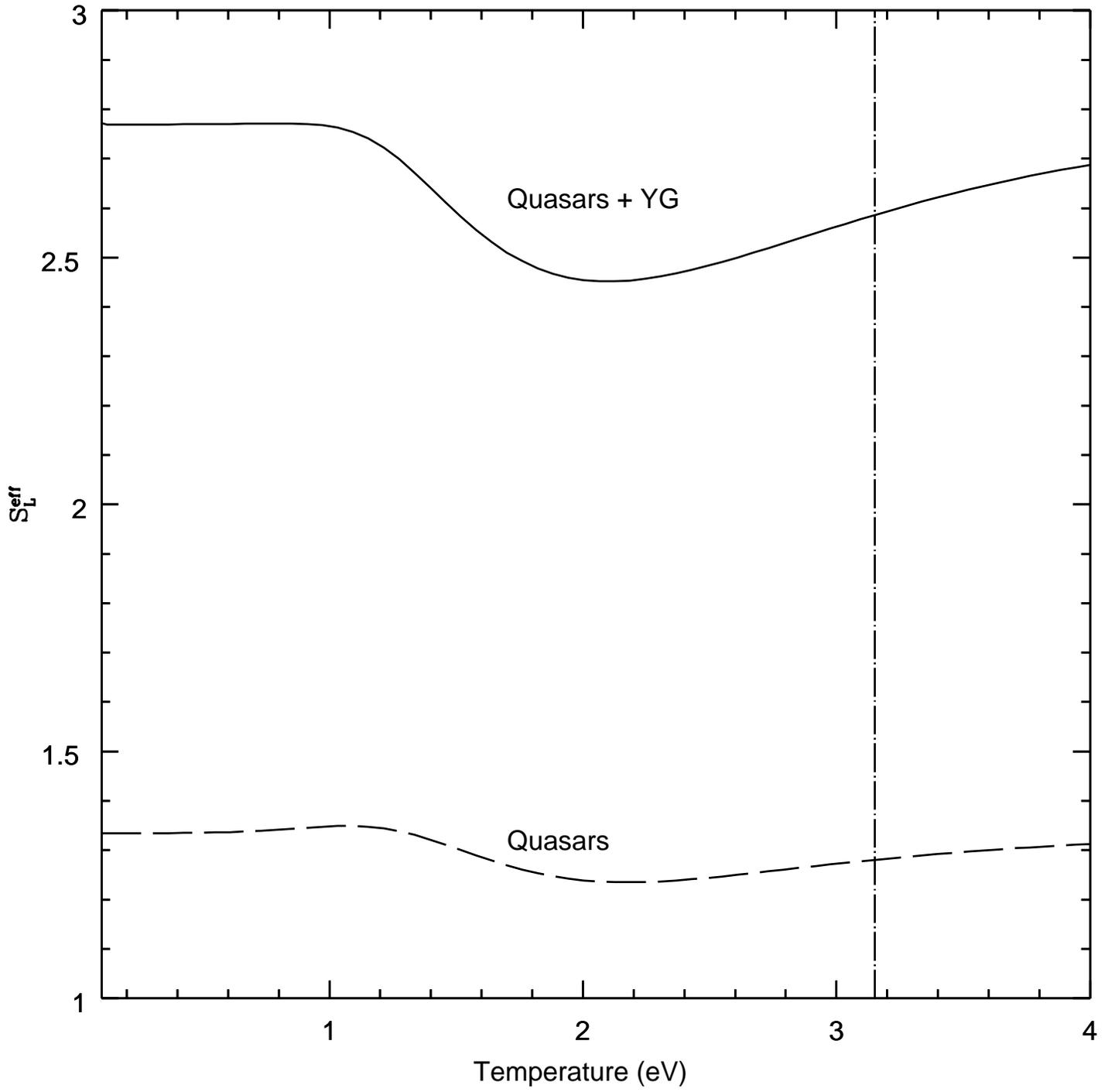
{\bf Figure 5} : The effective value of $S_L$ is plotted as a function
of the temperature of the IGM for two cases. The dashed curve is for
the case where only UV photons from quasars are considered. The solid curve
is for UV photons from quasars and young galaxies, where $\j21=1$ (at
$z=2.4$). Both curves use the absorption model A2. The dash-and-dotted
line shows the upper limit from the Compton y parameter bound.

\bs
\noindent
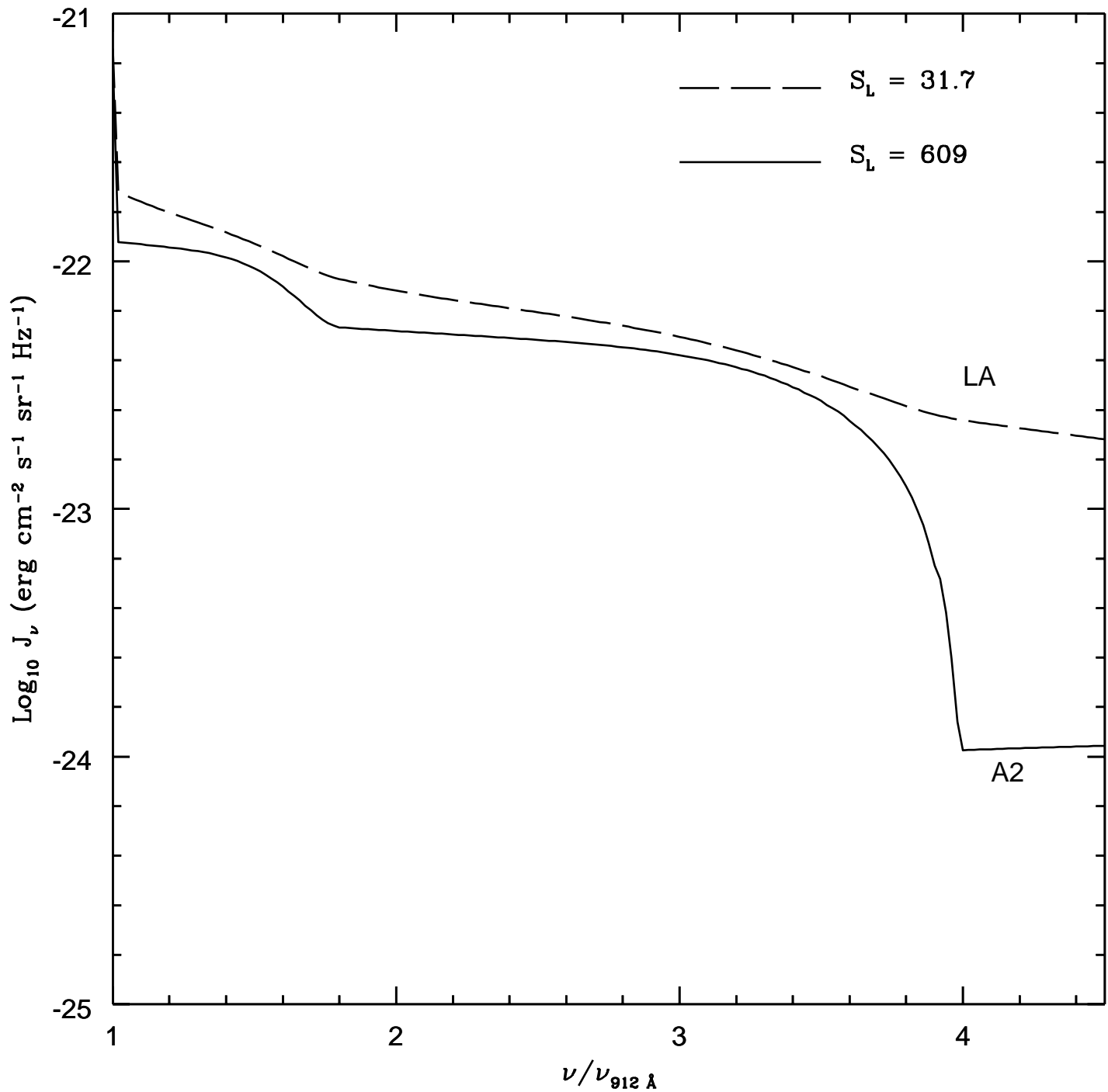
{\bf Figure 6} : The spectrum of the UV background radiation from quasars
and decaying neutrino is shown for two absorption models A2 and LA.

\noindent
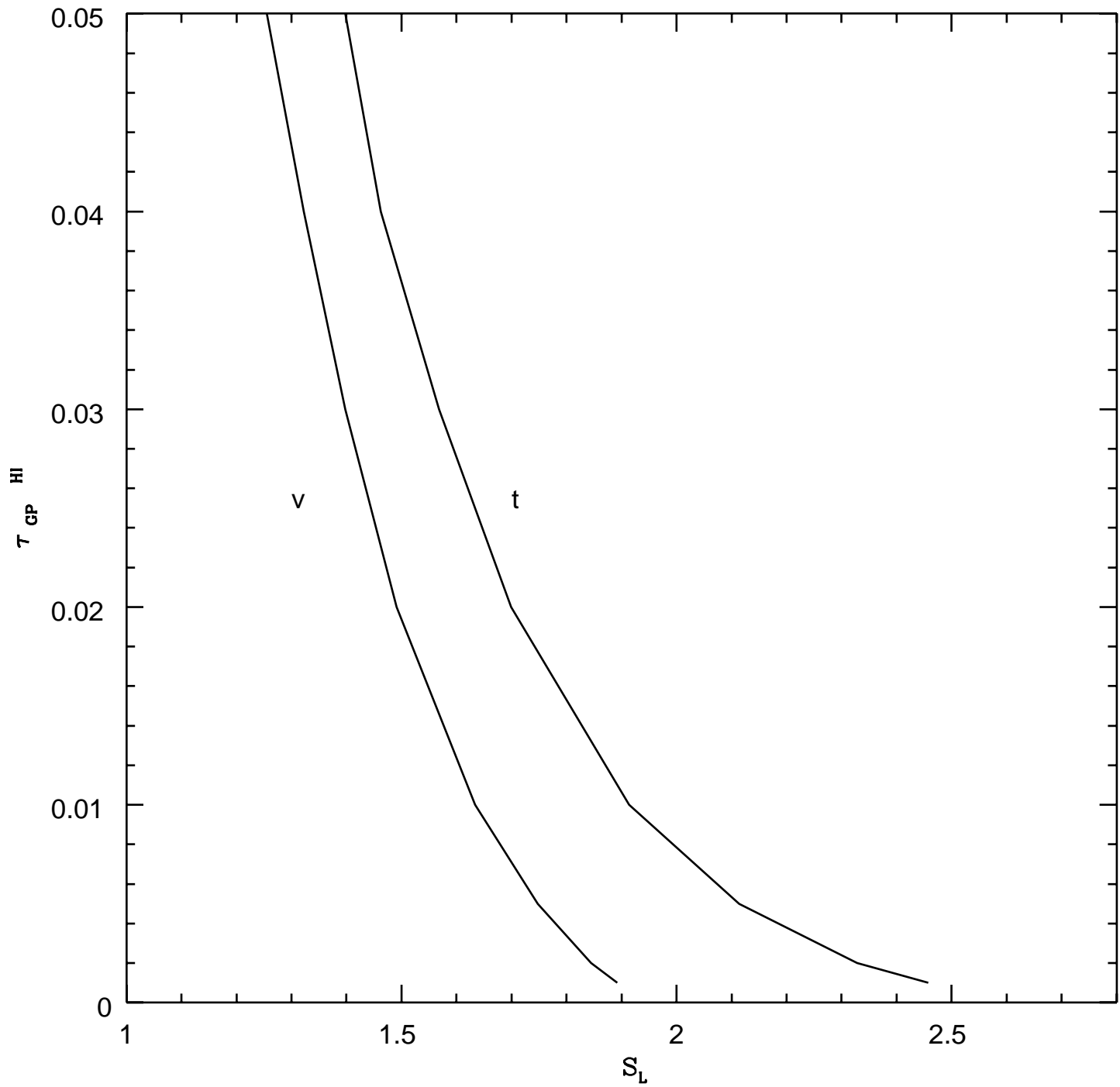
{\bf Figure 7}: The value of $\tauhgp$ that is necessary to explain
$\tau _{\scriptscriptstyle \rm He II} (z=2.4)=1$, where the He~II opacity
includes the contribution from line blanketing, is shown as a function
of $S_L$. Curves due to velocity and thermally broadened lines are labeled
`v' and `t' respectively.

\newpage

\section{Appendix}

The specific intensity (in units of $\rm ergs \, cm^{-2} \, s^{-1}\, Hz^{-1}\,sr^{-1}$ ) of background photons observed at a frequency $\nu_0$ 
and redshift $z_0$ is given by

$$J(\nu_0, z_0) = {c H_0^{-1}\over 4 \pi} \int^\infty_{z_0} dz {(1 + z_0)^3 
\over (1+z)^5 (1+\Omega_0z)^{1/2}}\epsilon(\nu,z)\exp[-\tau(\nu_{\rm obs},z_{\rm obs},z)]$$
where $\epsilon(\nu,z)$ is the  cumulative proper volume emissivity (in units of $\rm ergs \, cm^{-3} \, s^{-1}\, Hz^{-1}$)  of all the sources of 
photoionization (QSO, young galaxy, decaying neutrino, etc) at frequency $\nu = \nu_{\rm obs} (1+z)/(1+z_{\rm obs})$ and redshift $z$. $\tau(\nu_{\rm obs},z_{\rm obs},z)$ measure the optical depth suffered by a photon emitted at  redshift
$z$ and observed at redshift $z_{\rm obs}$ with frequency $\nu_{\rm obs}$. 

We consider two sources of absorption: diffuse IGM 
and Lyman-$\alpha$ systems. Therefore,
\begin{equation}
\tau(\nu_{\rm obs},z_{\rm obs},z) = \tau_{\sc \rm diffuse}(\nu_{\rm obs},z_{\rm obs},z) + \tau_{\rm cloud}(\nu_{\rm obs},z_{\rm obs},z)
\end{equation}
 Optical depth due to diffuse IGM is
\begin{equation}
  \tau_{\sc \rm diffuse}(\nu_{\rm obs},z_{\rm obs},z) \simeq \int^{z}_{z_{obs} }dz 
\left \vert{dt \over dz} \right\vert \, \left [n_{\sc\rm H}
\sigma_{\sc\rm HI}(\nu)y + n_{\sc\rm He}
\sigma_{\sc\rm HeI}(\nu)y_1 +  n_{\sc\rm He}
\sigma_{\sc\rm HeII}(\nu)y_2 \right ].
\end{equation}
Here, $y = n_{\sc\rm HI}/n_{\sc\rm H}$, $y_1 = n_{\sc\rm HeI}/n_{\sc\rm He}$, and $y_2 = n_{\sc\rm HeII}/n_{\sc\rm 
He}$.
$\tau_{\rm cloud}$, the average attenuation of photon 
flux due to poisson-distributed clouds, is given by 
\begin{eqnarray}
\tau_{\rm cloud}(\nu_{\rm obs},z_{\rm obs},z) & \simeq & \int_{z_{obs}}^z \int_0^\infty 
dz \, dN_{\sc\rm HI}\, {\cal P}(N_{\sc \rm HI},z) \nonumber \\
& &{} \times \{1-\exp 
\bigl \lbrack  - [ N_{\sc\rm HI}\sigma_{\sc\rm HI}(\nu) +
N_{\sc\rm HeI}\sigma_{\sc\rm HeI}(\nu) + N_{\sc\rm HeII}\sigma_{\sc\rm HeII}(\nu)] \bigr \rbrack\}. 
\end{eqnarray}
Here $N_{\sc\rm HI}$,  $N_{\sc\rm HeI}$, and $N_{\sc\rm HeII}$ correspond to column 
densities of neutral hydrogen, neutral helium, and singly ionized helium in the 
clouds and  ${\cal P}(N_{\sc\rm HI},z)$ is the  observed column density
and redshift distribution of the $\lyal$ clouds along an average
line of sight. In
 ionization equilibrium, $N_{\sc\rm HeI}$ and $N_{\sc\rm HeII}$ can be inferred from 
known column densities of neutral hydrogen $N_{\sc\rm HI}$  
\begin{equation}
 N_{\sc\rm HeII} = N_{\sc\rm HI} \times 1.8{ J_{912} \over J_{228}} \quad {\rm cm^{-2}} \quad \hbox{and} \quad N_{\sc\rm HeI} = N_{\sc\rm HI} \times 0.044{ J_{912} \over J_{504}} y_2 \quad \rm cm^{-2}. 
\end{equation}

The absorption due to neutral helium is important only for the case of
radiatively decaying neutrinos, as seen in Fig. 6. We consider two models A2  of MO (the same as  MA of Meiksin and Madau (1993) and model 3 of
Giroux and Shapiro (1996) and LA of Meiksin and Madau (1993) of absorption due
to $\lyal$ systems. Most realistic models of absorption due to $\lyal$ systems
fall between these two models (for a detailed discussion see Giroux and Shapiro (1996)).

The rates of photoionization, collisional ionization  and recombination 
coefficients are taken from 
Cen (1992).

\end{document}